\documentclass[ALICE,manyauthors]{cernphprep}
\usepackage[comma,square,numbers,sort&compress]{natbib}
\usepackage{hyperref}
\usepackage{comment}
\usepackage{lineno}
\usepackage{xspace}
\usepackage{color}

\usepackage[T1]{fontenc}
\usepackage{orcidlink}

\begin{document}
%

\newcommand{\pp}           {pp\xspace}
\newcommand{\ppbar}        {\mbox{$\mathrm {p\overline{p}}$}\xspace}
\newcommand{\XeXe}         {\mbox{Xe--Xe}\xspace}
\newcommand{\PbPb}         {\mbox{Pb--Pb}\xspace}
\newcommand{\pA}           {\mbox{pA}\xspace}
\newcommand{\pPb}          {\mbox{p--Pb}\xspace}
\newcommand{\AuAu}         {\mbox{Au--Au}\xspace}
\newcommand{\dAu}          {\mbox{d--Au}\xspace}

\newcommand{\s}            {\ensuremath{\sqrt{s}}\xspace}
\newcommand{\snn}          {\ensuremath{\sqrt{s_{\mathrm{NN}}}}\xspace}
\newcommand{\pt}           {\ensuremath{p_{\rm T}}\xspace}
\newcommand{\meanpt}       {$\langle p_{\mathrm{T}}\rangle$\xspace}
\newcommand{\ycms}         {\ensuremath{y_{\rm CMS}}\xspace}
\newcommand{\ylab}         {\ensuremath{y_{\rm lab}}\xspace}
\newcommand{\etarange}[1]  {\mbox{$\left | \eta \right |~<~#1$}}
\newcommand{\yrange}[1]    {\mbox{$\left | y \right |~<~#1$}}
\newcommand{\dndy}         {\ensuremath{\mathrm{d}N_\mathrm{ch}/\mathrm{d}y}\xspace}
\newcommand{\dndeta}       {\ensuremath{\mathrm{d}N_\mathrm{ch}/\mathrm{d}\eta}\xspace}
\newcommand{\avdndeta}     {\ensuremath{\langle\dndeta\rangle}\xspace}
\newcommand{\dNdy}         {\ensuremath{\mathrm{d}N_\mathrm{ch}/\mathrm{d}y}\xspace}
\newcommand{\Npart}        {\ensuremath{N_\mathrm{part}}\xspace}
\newcommand{\Ncoll}        {\ensuremath{N_\mathrm{coll}}\xspace}
\newcommand{\dEdx}         {\ensuremath{\textrm{d}E/\textrm{d}x}\xspace}
\newcommand{\RpPb}         {\ensuremath{R_{\rm pPb}}\xspace}

\newcommand{\nineH}        {$\sqrt{s}~=~0.9$~Te\kern-.1emV\xspace}
\newcommand{\seven}        {$\sqrt{s}~=~7$~Te\kern-.1emV\xspace}
\newcommand{\twoH}         {$\sqrt{s}~=~0.2$~Te\kern-.1emV\xspace}
\newcommand{\twosevensix}  {$\sqrt{s}~=~2.76$~Te\kern-.1emV\xspace}
\newcommand{\five}         {$\sqrt{s}~=~5.02$~Te\kern-.1emV\xspace}
\newcommand{\twosevensixnn}{$\sqrt{s_{\mathrm{NN}}}~=~2.76$~Te\kern-.1emV\xspace}
\newcommand{\fivenn}       {$\sqrt{s_{\mathrm{NN}}}~=~5.02$~Te\kern-.1emV\xspace}
\newcommand{\LT}           {L{\'e}vy-Tsallis\xspace}
\newcommand{\GeVc}         {Ge\kern-.1emV/$c$\xspace}
\newcommand{\MeVc}         {Me\kern-.1emV/$c$\xspace}
\newcommand{\TeV}          {Te\kern-.1emV\xspace}
\newcommand{\GeV}          {Ge\kern-.1emV\xspace}
\newcommand{\MeV}          {Me\kern-.1emV\xspace}
\newcommand{\GeVmass}      {Ge\kern-.2emV/$c^2$\xspace}
\newcommand{\MeVmass}      {Me\kern-.2emV/$c^2$\xspace}
\newcommand{\lumi}         {\ensuremath{\mathcal{L}}\xspace}

\newcommand{\ITS}          {\rm{ITS}\xspace}
\newcommand{\TOF}          {\rm{TOF}\xspace}
\newcommand{\ZDC}          {\rm{ZDC}\xspace}
\newcommand{\ZDCs}         {\rm{ZDCs}\xspace}
\newcommand{\ZNA}          {\rm{ZNA}\xspace}
\newcommand{\ZNC}          {\rm{ZNC}\xspace}
\newcommand{\SPD}          {\rm{SPD}\xspace}
\newcommand{\SDD}          {\rm{SDD}\xspace}
\newcommand{\SSD}          {\rm{SSD}\xspace}
\newcommand{\TPC}          {\rm{TPC}\xspace}
\newcommand{\TRD}          {\rm{TRD}\xspace}
\newcommand{\VZERO}        {\rm{V0}\xspace}
\newcommand{\VZEROA}       {\rm{V0A}\xspace}
\newcommand{\VZEROC}       {\rm{V0C}\xspace}
\newcommand{\Vdecay} 	   {\ensuremath{V^{0}}\xspace}

\newcommand{\ee}           {\ensuremath{e^{+}e^{-}}} 
\newcommand{\pip}          {\ensuremath{\pi^{+}}\xspace}
\newcommand{\pim}          {\ensuremath{\pi^{-}}\xspace}
\newcommand{\kap}          {\ensuremath{\rm{K}^{+}}\xspace}
\newcommand{\kam}          {\ensuremath{\rm{K}^{-}}\xspace}
\newcommand{\pbar}         {\ensuremath{\rm\overline{p}}\xspace}
\newcommand{\kzero}        {\ensuremath{{\rm K}^{0}_{\rm{S}}}\xspace}
\newcommand{\lmb}          {\ensuremath{\Lambda}\xspace}
\newcommand{\almb}         {\ensuremath{\overline{\Lambda}}\xspace}
\newcommand{\Om}           {\ensuremath{\Omega^-}\xspace}
\newcommand{\Mo}           {\ensuremath{\overline{\Omega}^+}\xspace}
\newcommand{\X}            {\ensuremath{\Xi^-}\xspace}
\newcommand{\Ix}           {\ensuremath{\overline{\Xi}^+}\xspace}
\newcommand{\Xis}          {\ensuremath{\Xi^{\pm}}\xspace}
\newcommand{\Oms}          {\ensuremath{\Omega^{\pm}}\xspace}
\newcommand{\degree}       {\ensuremath{^{\rm o}}\xspace}

\newcommand{\rhovnpt}{\ensuremath{\rho(v_n^2,[\pt])}\xspace}
\newcommand{\mpt}{\ensuremath{[\pt]}\xspace}
\newcommand{\vn}{\ensuremath{v_n}\xspace}
\newcommand{\vtwo}{\ensuremath{v_2}\xspace}
\newcommand{\vtwotwo}{\ensuremath{v_2\lbrace 2 \rbrace}\xspace}
\newcommand{\cov}{\ensuremath{\mathrm{cov}}\xspace}
\newcommand{\var}{\ensuremath{\mathrm{var}}\xspace}
\newcommand{\alert}[1]{\textcolor{red}{#1}}
\newcommand{\gevc}{\ensuremath{\mathrm{GeV}/c}\xspace}
\newcommand{\Nch}{\ensuremath{N_\mathrm{ch}}\xspace}

\newcommand{\expval}[1]{\ensuremath{\mathbb{E}\left\lbrace #1 \right\rbrace}\xspace}
\newcommand{\trueval}[1]{\ensuremath{\left\langle #1 \right\rangle}\xspace}
\newcommand{\estval}[1]{\ensuremath{\left\langle #1 \right\rangle}\xspace}
\newcommand{\lr}[1]{\ensuremath{\left\langle\left\langle #1 \right\rangle\right\rangle}\xspace}

\newcommand{\ccomment}[1]{\textcolor{red}{#1}}
\newcommand{\bcomment}[1]{\textcolor{blue}{#1}}
\newcommand{\gcomment}[1]{\textcolor{green}{#1}}

\newcommand{\ipgmusic}{{IP-Glasma + MUSIC + UrQMD}\xspace}
\newcommand{\pythia}{{PYTHIA}\xspace}
\newcommand{\ampt}{{AMPT}\xspace}

\newcommand{\vman}{manuscript}
\newcommand{\vfig}{Fig.}
\newcommand{\vref}{Ref.}
\newcommand{\vtab}{Tab.}


\begin{titlepage}
\PHyear{2026}       
\PHnumber{056}      
\PHdate{04 Mar}  

\title{Measurement of correlations between elliptic flow and mean transverse momentum in pp, p--Pb, and Pb--Pb collisions at the LHC}
\ShortTitle{Flow and mean transverse momentum correlations in small systems}   

\Collaboration{ALICE Collaboration\thanks{See Appendix~\ref{app:collab} for the list of collaboration members}}
\ShortAuthor{ALICE Collaboration} 

\begin{abstract}
Measurements of the event-by-event correlation between elliptic flow ($v_2$) and the mean transverse momentum~($[\pt]$) using the modified Pearson correlation coefficient $\rho(v_2^2,[\pt])$ are reported in pp collisions at $\sqrt{s} = 13$~TeV, and in p--Pb and Pb--Pb collisions at a center-of-mass energy per nucleon pair $\sqrt{s_{_\mathrm{NN}}} = 5.02$~TeV.
This analysis is based on the full LHC Run 2 dataset recorded by ALICE and is performed for the first time in small collision systems with the ALICE detector.
In Pb--Pb collisions, the $\rho(v_2^2,[\pt])$ measurement shows a non-monotonic dependence on charged particle multiplicity ($\Nch$); it first decreases and then increases with an increase in multiplicity. The decreasing trends of $\rho(v_2^2,[\pt])$ with increasing multiplicity are also observed in p--Pb and pp collisions for the presented multiplicity range. All three systems show consistent values of $\rho(v_2^2,[\pt])$ for $\Nch \lesssim 80$. These measurements are also compared with theoretical model calculations, including PYTHIA, where no collectivity is generated, as well as AMPT and \ipgmusic, which produce collective effects in small systems. These comparisons offer unique insights into the origin of collectivity in small systems.
They improve the understanding of the initial geometry, size, and their correlations. The comparison also allows an investigation of the role of initial momentum correlations predicted by the Color Glass Condensate framework.
The new measurements could not be explained by current state-of-the-art models, offering insights into the initial stage of collisions in small systems and also imposing strong constraints on the existing theoretical models. This will significantly advance our understanding of the collective phenomena observed in small systems at the LHC.

\end{abstract}
\end{titlepage}

\setcounter{page}{2} 


\section{Introduction}
\label{sec:introcution}
The quark--gluon plasma (QGP), a deconfined state of quarks and gluons, is expected to be created in ultra-relativistic heavy-ion collisions~\cite{Collins:1974ky,Karsch:2001vs}. 
At the onset of the RHIC and LHC heavy-ion programs, small systems, such as proton-proton (pp) and proton-nucleus (p--A) collisions,
were considered baselines for heavy-ion studies, under the assumption that QGP formation does not occur in these systems~\cite{Braun-Munzinger:2007edi}.
However, patterns characteristic of collective effects, typically associated with QGP formation in high-energy heavy-ion collisions, have been observed in these small systems at RHIC~\cite{PHENIX:2014fnc,STAR:2015kak,PHENIX:2018lia,STAR:2022pfn} and the LHC~\cite{CMS:2010ifv,CMS:2012qk,ATLAS:2012cix,ALICE:2012eyl,ALICE:2013snk,ALICE:2019zfl,ALICE:2024vzv}.
The first experimental observation of collective-like effects (i.e., exhibiting signatures similar to collective effects observed in heavy-ion collisions) in small collision systems was the discovery of a near-side ridge structure~\cite{CMS:2010ifv}. This structure represents long-range azimuthal correlations between produced particles, extending across a large pseudorapidity range in pp collisions. 
Following this discovery, numerous studies were conducted to further investigate the presence of collective-like effects in small collision systems.
These include measurements of azimuthal anisotropies~\cite{ATLAS:2012cix,ALICE:2012eyl,ATLAS:2014qaj,PHENIX:2014fnc,ATLAS:2015hzw,CMS:2015fgy,STAR:2015kak,CMS:2016fnw,CMS:2017kcs,ATLAS:2017hap,PHENIX:2018lia,ALICE:2019zfl,STAR:2022pfn,ALICE:2023gyf}, multi-particle correlations~\cite{CMS:2012qk, CMS:2016fnw,ATLAS:2017hap,ALICE:2019zfl}, and anisotropic flow using identified hadrons~\cite{ALICE:2013snk,CMS:2014und,CMS:2018loe,CMS:2022bmk,ALICE:2024vzv}.
Recent comprehensive reviews of these measurements can be found in Refs.~\cite{Grosse-Oetringhaus:2024bwr,Christiansen:2024bhe}.

The possibility of collective effects in small collision systems raises essential questions about their underlying mechanisms. 
Most theoretical models describe a pp collision as a sequence of distinct stages: the initial conditions, dynamical evolution, and hadronization into final-state particles.
Different models incorporate distinct physical mechanisms at each stage to reproduce the observed features in the final state.
At the initial stage, spatial eccentricity may arise from fluctuations in the parton distribution within the proton~\cite{Schenke:2012wb,Schenke:2018fci} or due to overlapping strings~\cite{Lin:2004en}. 
Additionally, models based on the Color Glass Condensate (CGC) predict an initial momentum anisotropy, i.e., the momentum correlations of partons in the initial stage of collisions~\cite{Schenke:2012wb,Schenke:2018fci}.
The system's evolution can be modeled in different ways. Relativistic hydrodynamic models~\cite{Schenke:2010nt,Song:2007ux,Song:2007fn} describe the evolution of the matter created in the high-energy collisions by treating it as a relativistic fluid, using pressure gradients and viscosity to describe its expansion and cooling. 
In contrast, parton transport models~\cite{Lin:2004en,Xu:2007aa} simulate quasi-particle interactions, offering a microscopic perspective on system dynamics. The system then experiences hadronization and resonance decay, resulting in the emission of final-state hadrons.
Most observables, such as anisotropic flow, are sensitive to initial effects and the system evolutions~\cite{Schenke:2020mbo,Zhao:2021bef}.
As a result, it is difficult to use such observables to disentangle the contributions from the initial and final stages.
This poses a challenge when attempting to test specific mechanisms or constrain model parameters.
Therefore, it is important to find an observable that is sensitive only to the initial-stage effects and less affected by the complex dynamic evolution of the created matter.

Measurements of correlations between the strength of anisotropic flow and the mean transverse momentum based on the modified Pearson coefficient $\rho(v_2^2, [p_{\rm T}])$~\cite{Bozek:2016yoj} have been identified as a promising probe of fluctuations occurring during the initial stage of collisions~\cite{Giacalone:2020dln,Zhao:2024lpc}.
The $\rho(v_2^2, [p_{\rm T}])$ observable is defined as
\begin{equation}
   \rho(v_2^2, [p_{\rm T}]) = 
   \frac{\mathrm{cov}(v_2^2, [p_{\rm T}])}{\sqrt{\mathrm{var}(v_2^2)} \sqrt{\mathrm{var}([p_{\rm T}]})}.
   \label{eq:rho}
\end{equation}
The $\cov(v_2^2,[\pt])$ is the covariance between the event-wise squared elliptic flow coefficient, $v_2^2$, and the event-by-event mean transverse momentum, $[\pt]$. It is calculated as $\langle v_2^2\delta\pt \rangle$,
where $\delta\pt = [\pt] - \langle[\pt]\rangle$ is the 
deviation of each event's mean $\pt$ from the average over all events in a given multiplicity bin.
The value of $v_2^2$ is obtained from the two-particle correlations~\cite{Borghini:2000sa}. The $\var(v_2^2)$ is the dynamic variance of the elliptic flow and is defined as $\var(v_2^2) = \langle{v_2^4}\rangle - \langle{v_2^2}\rangle^2$,
and $\var([\pt]) = \langle{\delta\pt^2}\rangle$ is the variance of $[\pt]$, measuring the event-by-event fluctuations of the mean transverse momentum.
The angular brackets $\langle...\rangle$ denote an average computed over many events. 
In non-central heavy-ion collisions, fluctuations in $[\pt]$ primarily reflect variations in the size of the created fireball~\cite{Broniowski:2009fm}, while $v_2^2$ is mainly governed by the initial geometrical eccentricity $\varepsilon_2^2$~\cite{Alver:2010dn,Niemi:2012aj}. Thus, the measurement of $\rho(v_2^2,[\pt])$ in heavy-ion collisions serves as a probe for initial-state fluctuations, characterized by the correlation between the initial geometrical eccentricity $\varepsilon_2^2$ and the size of the created fireball~\cite{Giacalone:2020dln,Schenke:2020uqq,Lim:2021auv,Zhao:2024lpc,ALICE:2021gxt}. 
With the additional effects predicted by the CGC models, it was shown in the theoretical model calculations that in pp and p--Pb collisions, the coefficient $\rho(v_2^2, [p_{\rm T}])$ changes from positive values at low multiplicity to negative values at high multiplicity~\cite{Giacalone:2020byk}.
For Pb--Pb collisions, as multiplicity increases, the coefficient $\rho(v_2^2, [p_{\rm T}])$ initially changes from positive to negative and then returns to positive at high multiplicity, which is explained in Ref.~\cite{Giacalone:2020byk} as a combined effect of CGC and geometric influences.
A single sign change in pp and p--Pb collisions, and a double sign change in Pb--Pb collisions, have been proposed as signatures of the CGC~\cite{Giacalone:2020byk}.
Observables such as anisotropic flow coefficients exhibit similar sensitivities to both the CGC effect and the geometry-seeded hydrodynamical response~\cite{Schenke:2020mbo}, making it difficult to determine the origin of flow in these studies. 
In contrast, $\rho(v_2^2, [p_{\rm T}])$ offers unique discriminatory power, as CGC and geometry-driven contributions yield opposite signs and different multiplicity dependencies, particularly in low-multiplicity events~\cite{Giacalone:2020byk}.
Therefore, measuring $\rho(v_2^2, [p_{\rm T}])$ and comparing it with theoretical models can improve our understanding of the initial-state properties in small system collisions, including shape, size, and the possible contribution from the CGC.

The Pearson coefficient $\rho(v_2^2, [p_{\rm T}])$ was previously measured in Pb--Pb collisions at the LHC~\cite{ALICE:2021gxt, ATLAS:2022dov}, where the results demonstrated unique sensitivity to the initial conditions, particularly the nucleon size, which had not been well constrained before~\cite{Giacalone:2021clp}. It was found that both the measurements and theoretical model calculations strongly depend on the choices of centrality determination, whether based on particle production at midrapidity or forward rapidities. Since most popular hydrodynamic calculations are performed using a 2+1D hydrodynamic framework, they face challenges in reliably simulating particle production in the forward region, as used in previous studies~\cite{ALICE:2021gxt}. To avoid this issue, the present study performs the $\rho(v_2^2, [p_{\rm T}])$ measurements as a function of charged-particle multiplicity at midrapidity.
In addition, the $\rho(v_2^2, [p_{\rm T}])$ observable has also been measured in small collision systems in the range of $0.3 < \pt < 2.0~\mathrm{GeV}/c$~\cite{ATLAS:2019pvn} or $0.5 < \pt < 5.0~\mathrm{GeV}/c$~\cite{CMS:2024rvk}.
As suggested in Ref.~\cite{Zhang:2021phk}, excluding particles with high transverse momentum reduces contamination by hard jets, which can bring significant nonflow contaminations and dilute sensitivity to the initial stages of the collision. Additionally, it was shown that the initial momentum anisotropy originating from CGC reduces rapidly with increasing pseudorapidity separations~\cite{Schenke:2022mjv}.
This implies that the previous measurements may be less sensitive to contributions from the CGC.
To maintain sensitivity to the CGC while suppressing nonflow effects, measurements should be conducted at lower transverse momentum and within a narrower pseudorapidity gap between the particles to measure the observable, which is achievable with the ALICE detector.

\section{Experimental setup and analysis details}
\label{sec:cuts}
This analysis uses the full datasets for pp collisions at $\sqrt{s} = 13~\mathrm{TeV}$, and p--Pb and Pb--Pb collisions at $\sqrt{s_\mathrm{NN}} = 5.02~\mathrm{TeV}$, recorded by ALICE during the LHC Run 2 period~\cite{ALICE:2008ngc,ALICE:2014sbx}.

Event and track selection criteria are applied to ensure a high-quality dataset.
This selection process includes both online trigger selection and offline selection of events.
In p--Pb and Pb--Pb collisions, a minimum bias (MB) trigger is employed, based on time-coincident signals between the V0A and V0C detectors~\cite{ALICE:2013axi}, which are segmented scintillators covering the full azimuth within the pseudorapidity ranges $2.8 < \eta < 5.1$ (V0A) and $-3.7 < \eta < -1.7$ (V0C).
In pp collisions, a high-multiplicity (HM) trigger is used to select events with significantly higher activity. 
Events are triggered when the V0M signal exceeds four to five times the MB average, depending on the data taken period.
The V0M amplitude is defined as the sum of the signals from V0A and V0C.
This effectively selects the top 0.1\% of MB events with the highest particle multiplicity measured by V0M.

Primary tracks are reconstructed based on hits from the Inner Tracking System (ITS)~\cite{ALICE:2010tia} and the Time Projection Chamber (TPC)~\cite{Alme:2010ke}.
The ITS consists of six cylindrical silicon layers surrounding the beam pipe, and the TPC reconstructs charged-particle tracks using up to 159 three-dimensional space points along their trajectories.
The accepted tracks must have a minimum of 70 TPC clusters out of 159 to ensure high track quality. Additionally, the maximum $\chi^2$ value per TPC cluster should not exceed 2.5.
At least one hit in the Silicon Pixel Detector (SPD) detector is required, and the maximum $\chi^2$ value per ITS cluster must not exceed 3.6. 
A criterion on the distance of the closest approach (DCA) to the primary vertex is applied to suppress contamination from tracks produced  in weak decays and secondary interactions within the detector.
The longitudinal DCA ($|\mathrm{DCA}_z|$), must be less than $2~\mathrm{cm}$. 
The transverse DCA ($\mathrm{DCA}_{xy}$), must satisfy $\mathrm{DCA}_{xy} < 7  \times(0.0026+0.005/p_{\mathrm{T}}^{1.01})~\mathrm{cm}$, with $p_\mathrm{T}$ in $\mathrm{GeV}/c$. After event selection, the numbers of events in pp, p–Pb, and Pb–Pb collisions are $8.8\times10^8$, $5.5\times10^8$, and $1.49\times10^8$, respectively.

The measurements are carried out using the subevent method following the same configuration as in Ref.~\cite{ALICE:2021gxt}.
Specifically, the mean transverse momentum ($[\pt]$) is measured using particles taken from the subevent A where $|\eta| < 0.4$.
Likewise, the $v_2^2$ is extracted from two-particle correlations with a separation of $|\Delta\eta| > 0.8$, using particles from subevent B ($-0.8 < \eta < -0.4$) and subevent C ($0.4 < \eta < 0.8$). 
All the particles selected in the measurements are within the kinematic region of $0.2 < \pt < 3.0~\mathrm{GeV}/c$.

The systematic uncertainties related to event and track selection are evaluated by varying the selection criteria.
The systematic uncertainty associated with event selection of primary vertex position is estimated by altering the primary vertex position requirement along the beam direction, tightening it from $|z_\mathrm{vtx}| < 10~\mathrm{cm}$ to $|z_\mathrm{vtx}| < 7~\mathrm{cm}$.
One source of systematic uncertainty from track quality selection is investigated by including tracks without SPD hits.
This accounts for the absence of tracks with SPD hits in certain regions due to dead channels in the SPD.
An alternative selection criterion is increasing the minimum required number of TPC clusters to 80, yielding a subsample of tracks with improved reconstruction quality.
Additionally, to estimate the impact from the DCA selection, the longitudinal DCA limit is tightened to $|\mathrm{DCA}_z| < 1~\mathrm{cm}$ and the transverse DCA selection criteria to $\mathrm{DCA}_{xy} < 4\times(0.0026+0.005/p_{\mathrm{T}}^{1.01})~\mathrm{cm}$ with $p_\mathrm{T}$ in $\mathrm{GeV}/c$. These stricter selections reduce contributions from secondary particles and material interactions.
Each of these modifications was implemented individually to assess its impact. The absolute uncertainties of $\rho(v_2^2,[\pt])$ introduced by these sources are less than 0.005. 
To better understand the detector effect, a Monte Carlo closure test is performed. It compares the $\rho(v_2^2,[\pt])$ results obtained at the event generator level from AMPT~\cite{Lin:2004en} with the simulation output after the full reconstruction through the GEANT3 simulation~\cite{Brun:1994aa}. This MC study yields a systematic uncertainty of less than 0.03.
The statistically significant~\cite{Barlow:2002yb} sources of systematic uncertainty are added in quadrature to get the total absolute systematic uncertainty of less than 0.03, denoted as boxes in the figures in Sec.~\ref{sec:results}.

\section{Results}
\label{sec:results}
\label{sec:pcc}

\subsection{Results of \texorpdfstring{$\rho(v_2^2,[\pt])$}{rho(v2,pt)} in different collision systems}
\label{sec:re-1}

\begin{figure}[!hbt]
\centering
    \includegraphics[width=0.7\textwidth]{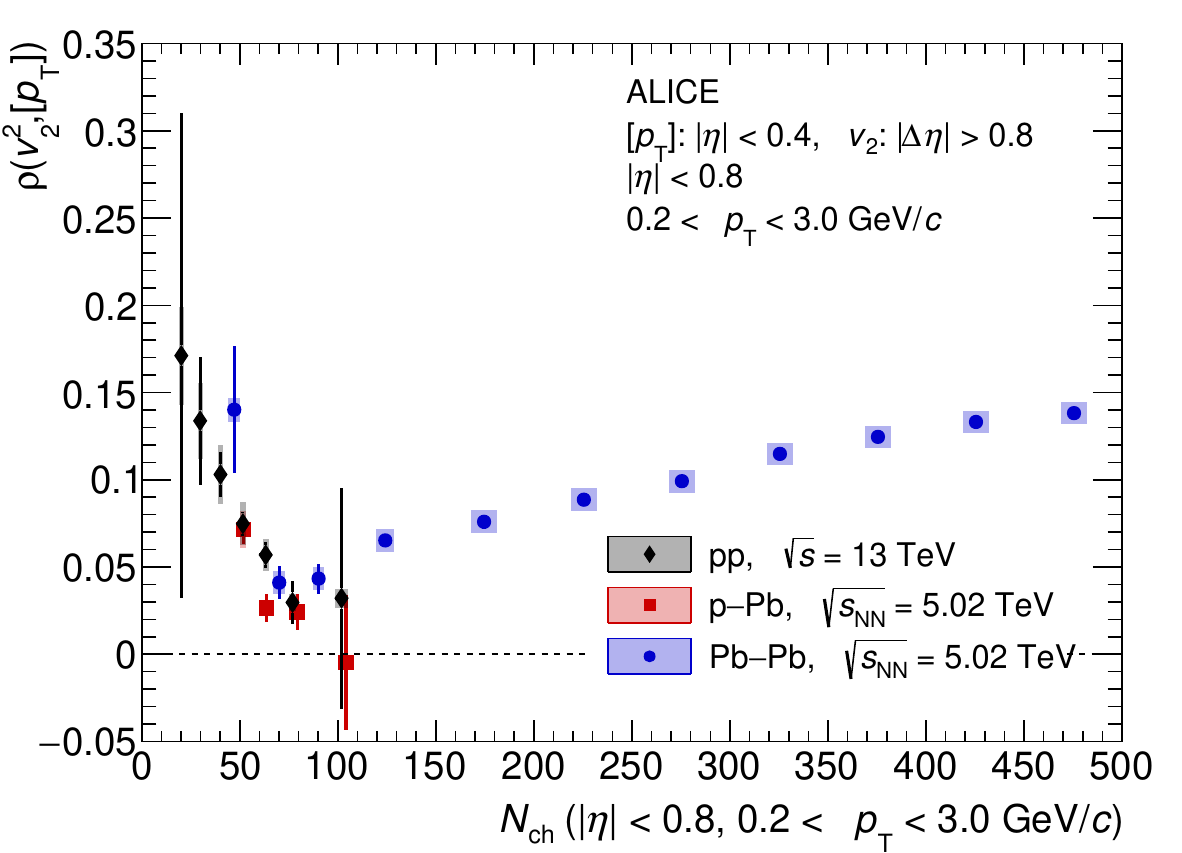}
    \caption{Measurements of $\rho(v_2^2,[\pt])$ as a function of charged particle multiplicity in pp, p--Pb and Pb--Pb collisions. Statistical uncertainties are shown as vertical lines, and systematic uncertainties as filled boxes.}
    \label{fig:rhovnpt}
\end{figure}

Values of $\rho(v_2^2,[\pt])$ measured as a function of charged particle multiplicity ($\Nch$) in pp, p--Pb, and Pb--Pb collisions, are shown in \vfig~\ref{fig:rhovnpt}.
The measured $\rho(v_2^2,[\pt])$ ranges from $-0.01$ to $0.2$, showing a weak correlation strength between $v_2^2$ and $[\pt]$.
In Pb--Pb collisions  $\rho(v_2^2,[\pt])$ decreases with decreasing multiplicity for $\Nch \gtrsim 80$ and then increases again for $\Nch \lesssim 80$.
Meanwhile, the measurements in pp and p--Pb collisions show clear increasing trends as multiplicity decreases.
For $\Nch \lesssim 80$, the measurements in all three collision systems are consistent within the uncertainties.
A possible explanation for the consistent measurements across the three collision systems at $\Nch \lesssim 80$ is that, in this low multiplicity region, the system size and shape are dominated by energy density fluctuations in the initial stage
~\cite{ALICE:2019zfl}.
As a result, the shape-size correlations and possible initial momentum correlations may be similar, regardless of system size.
The insensitivity of the results to the $|\Delta\eta|$ separation is supported by repeating the measurement with a larger $|\Delta\eta|$ separation in the  $v_2^2$ extraction, which yields results consistent with those in Fig.~\ref{fig:rhovnpt} (see Appendix~\ref{app:gap}). This indicates that the observed correlations are largely insensitive to non-collective (non-flow) effects.
To better interpret these results, comparisons with theoretical models incorporating different underlying mechanisms are essential.

\subsection{Comparison with model calculations}
\begin{figure}[!hbt]
\centering
    \includegraphics[width=0.7\textwidth]{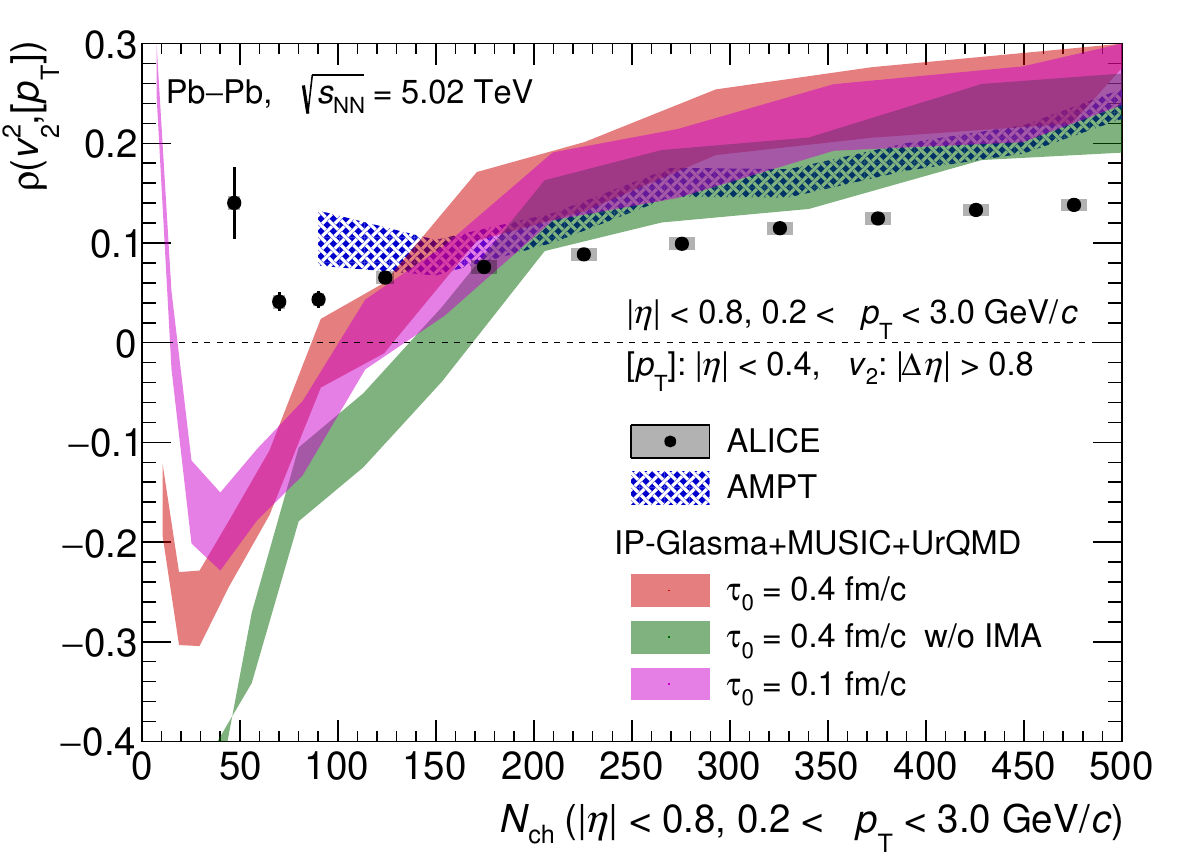}
    \caption{Measurements of $\rho(v_2^2,[\pt])$ as a function of charged particle multiplicity in Pb--Pb collisions, compared with the \ipgmusic~\cite{Giacalone:2020byk}, and \ampt~\cite{Lim:2021auv} calculations.}
    \label{fig:rhovnpt_model2PbPb}
\end{figure}

\begin{figure}[!hbt]
\centering
    \includegraphics[width=0.7\textwidth]{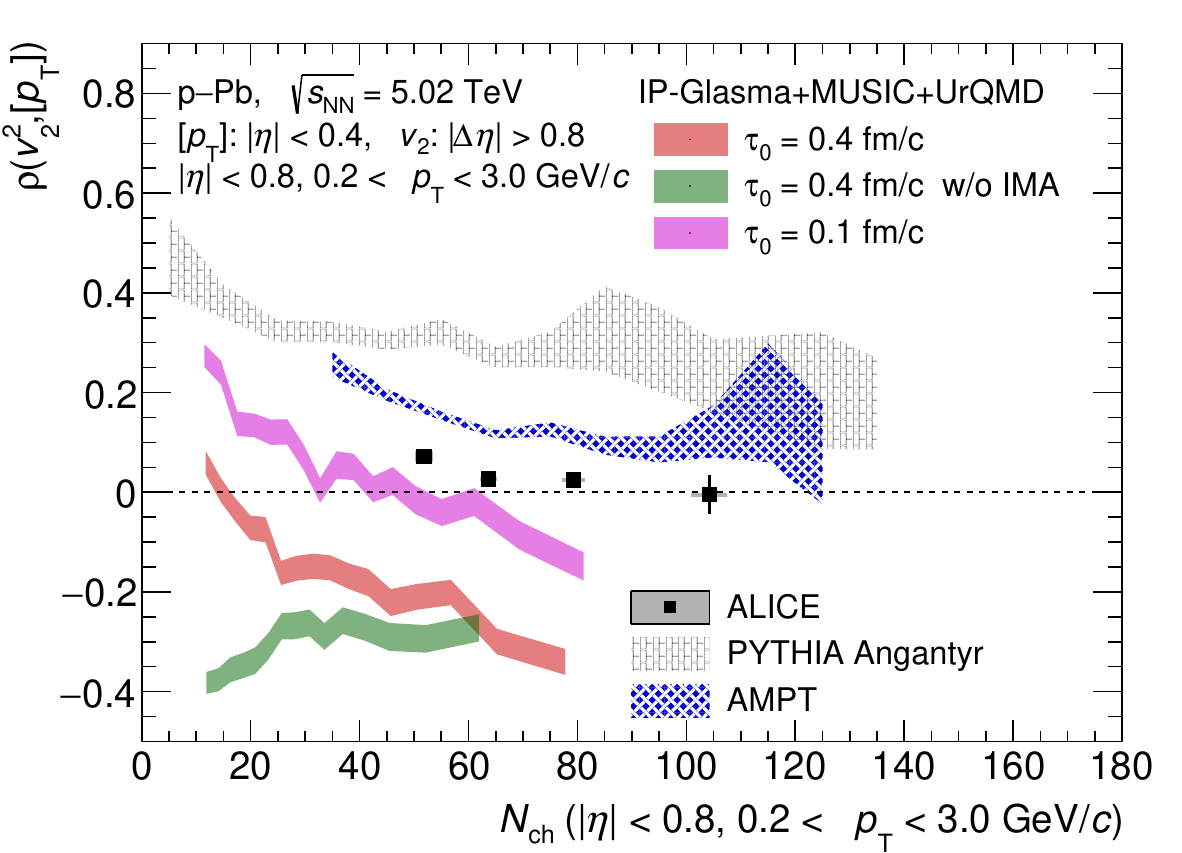}
    \caption{Measurements of $\rho(v_2^2,[\pt])$  as a function of charged particle multiplicity in p--Pb collisions, compared with the \ipgmusic~\cite{Giacalone:2020byk}, \ampt~\cite{Lim:2021auv} and PYTHIA~8~\cite{Lim:2021auv} calculations.}
    \label{fig:rhovnpt_model2pPb}
\end{figure}

\begin{figure}[!hbt]
\centering
    \includegraphics[width=0.7\textwidth]{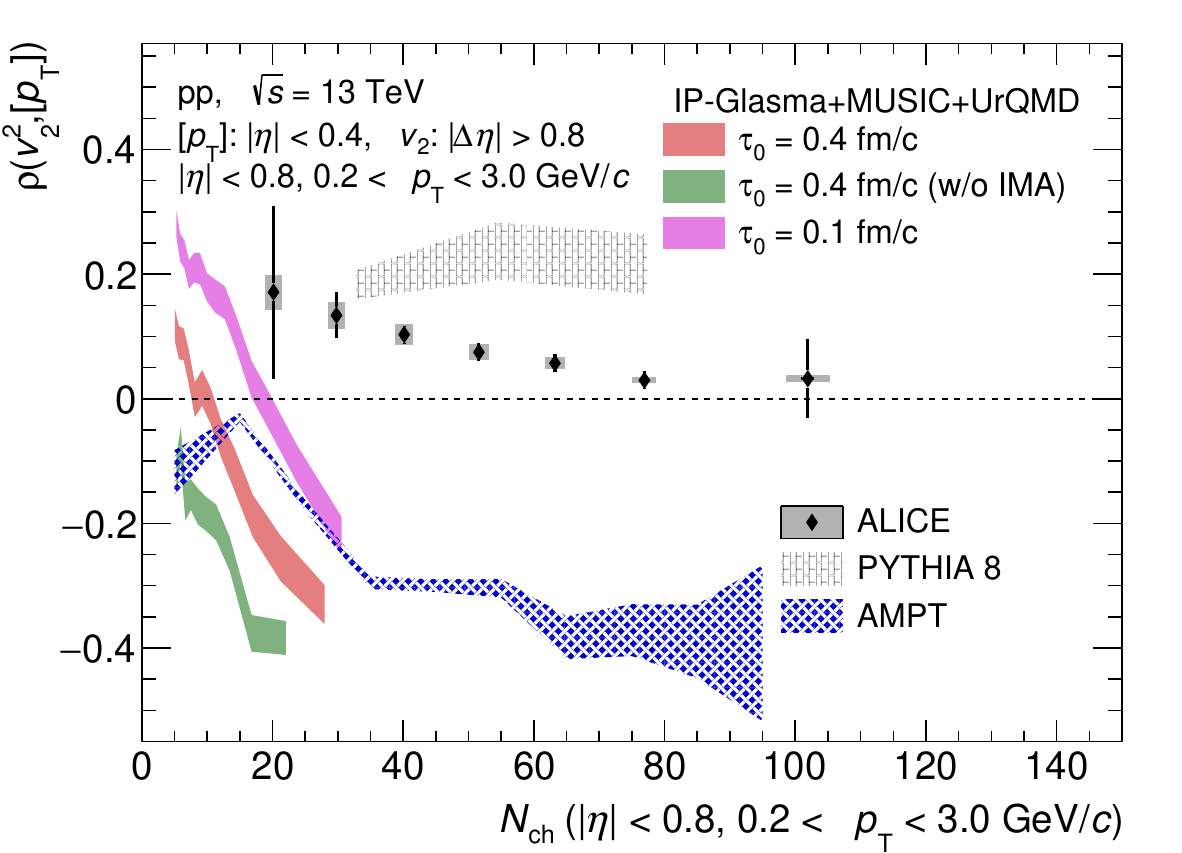}
    \caption{Measurements of $\rho(v_2^2,[\pt])$  as a function of charged particle multiplicity pp collisions, compared with the \ipgmusic~\cite{Giacalone:2020byk}, \ampt~\cite{Lim:2021auv} and PYTHIA~8~\cite{Lim:2021auv} calculations.}
    \label{fig:rhovnpt_model2pp}
\end{figure}

The $\rho(v_2^2,[\pt])$ measurements are compared with three different types of model calculations, \ipgmusic~\cite{Schenke:2012wb,Giacalone:2020byk}, \ampt~\cite{Lin:2004en,Lim:2021auv} and PYTHIA~\cite{Skands:2014pea,Bierlich:2018xfw,Lim:2021auv}, shown in Figs.~\ref{fig:rhovnpt_model2PbPb},~\ref{fig:rhovnpt_model2pPb}, and~\ref{fig:rhovnpt_model2pp}.
The model calculations are confined to regions where the generated sample is sufficiently large to ensure accuracy.
Specifically, the \ipgmusic employs the IP-Glasma for initial conditions, MUSIC for hydrodynamic evolution, and UrQMD for hadron rescattering and resonance decays~\cite{Schenke:2012wb,Giacalone:2020byk}. 
The \ipgmusic calculations are presented with and without the initial momentum anisotropy (IMA), which is produced by the CGC.
The \ipgmusic calculations with different initial times of hydrodynamic evolution ($\tau_0$) are also shown for the three collision systems.
Reducing $\tau_0$ from 0.4~$\mathrm{fm}/c$ to 0.1~$\mathrm{fm}/c$ effectively drives the system into a system with a strongly coupled initial stage, where the presence or absence of a glasma phase has little impact on the observables.
Moreover, the parton transport model AMPT has been shown to reproduce collective effects in small systems~\cite{He:2015hfa}.
The presented AMPT results~\cite{Lim:2021auv} use the string-melting version of the model~\cite{Lin:2004en}.
The PYTHIA~8 model~\cite{Skands:2014pea} is widely used for describing high-energy hadron collisions.
The PYTHIA Angantyr model~\cite{Bierlich:2018xfw} is an extended version of PYTHIA~8 designed to simulate hadron-nucleus and nucleus-nucleus collisions. Neither PYTHIA~8 nor Angantyr generates collective effects. Therefore, calculations using PYTHIA~8 and PYTHIA Angantyr~\cite{Lim:2021auv} and their comparisons to measurements can be used to investigate non-flow effects.

Figure~\ref{fig:rhovnpt_model2PbPb} presents a comparison of the measurement with the AMPT and \ipgmusic calculations in Pb--Pb collisions. The AMPT calculations show a mild decrease of in $\rho(v_2^2,[\pt])$ as \Nch decreases.
The AMPT results slightly overestimate the data but qualitatively match the observed trend in the range of $100 \lesssim \Nch \lesssim 500$, suggesting that its initial state shape-size correlations can reproduce the measured correlation of flow and mean transverse momentum in this region.
In the \ipgmusic calculations, $\rho(v_2^2,[\pt])$ decreases monotonically with decreasing multiplicity from $\Nch = $ 500 to 150, regardless of the hydrodynamic evolution time $\tau_0$ or the presence of IMA. Differences between model variants are minimal in this range.
The \ipgmusic calculations, in particular those without IMA, show good agreement with the AMPT calculations. As shown in Ref.~\cite{Lim:2021auv}, this agreement can be understood in terms of the correlation between the initial geometric size and shape of the overlap region in the transverse plane, which could be similar to these models.
For this multiplicity region, the \ipgmusic calculations with these variations qualitatively describe the $\rho(v_2^2,[\pt])$ measurements, but tend to slightly overestimate the observed values.
At the multiplicities range of $50 \lesssim \Nch \lesssim 150$, the \ipgmusic calculations predict a sign change from positive to negative as the multiplicity decreases, which is not observed in the measurements.
Moreover, the \ipgmusic calculation without the IMA deviates noticeably from the AMPT results for $50 \lesssim \Nch \lesssim 150$, suggesting that fluctuations lead to different shape–size correlation patterns in the two models.
At lower multiplicities ($\Nch \lesssim 50$), the default \ipgmusic model (with IMA and $\tau_0 = 0.4~\mathrm{fm}/c$) shows a rising trend, resulting in a non-monotonic behavior similar to ALICE data. However, unlike the model, the measurements remain positive throughout and do not exhibit a sign change.
Switching off the IMA from CGC causes the \ipgmusic to continue decreasing as $\Nch$ decreases, leading to a further deviation from the measurements. 
This highlights the crucial role of initial momentum anisotropy in reproducing the non-monotonic \Nch dependence seen in the measurements.
Varying the hydrodynamic start time $\tau_0$ value from 0.4 to $0.1~\mathrm{fm}/c$ shows no significant effect for  $\Nch \gtrsim 50$, but at lower multiplicity, the earlier start time yields a larger $\rho(v_2^2,[\pt])$ and reproduces the non-monotonic trend more closely to the measurements.

Figure~\ref{fig:rhovnpt_model2pPb} shows the comparison between measured $\rho(v_2^2,[\pt])$ and the calculations from AMPT, \ipgmusic and the PYTHIA Angantyr in p--Pb collisions.
As the multiplicity decreases, a slightly increasing trend is observed in the PYTHIA Angantyr calculations; the values are significantly larger than the $\rho(v_2^2,[\pt])$ measurements, which shows that non-flow effects can not fully explain the measurements. 
The AMPT calculation also shows a similar increasing trend with decreasing \Nch, and the calculation slightly overestimates the measurements. 
This general agreement suggests that in p--Pb collisions the shape-size correlations in AMPT's initial conditions can approximately reproduce the observed correlation of flow and mean transverse momentum.
A similar increasing trend is also seen in the default \ipgmusic calculation. 
However, for  $\Nch > 20$, this model predicts negative correlations, in contrast to the positive values observed in the measurements.
Using a smaller hydrodynamic starting time, $\tau_0 = 0.1~\mathrm{fm}/c$, which enhances early-stage interactions, leads to significantly larger $\rho(v_2^2,[\pt])$ values compared to the default $\tau_0 = 0.4~\mathrm{fm}/c$ setting. 
This shows the significant impact of strongly-coupled hydrodynamic effects in the early stage on the correlation between elliptic flow and mean transverse momentum.
A more detailed investigation of $\tau_0$ in small systems is needed, supported by studies of particle production and basic flow observables.
Moreover, removing the IMA from the \ipgmusic framework leads to significant deviations from the default calculation at $\Nch \lesssim 60$.
The IMA contribution not only modifies the \Nch-dependent trend but also reverses the sign of $\rho(v_2^2,[\pt])$ at very low \Nch. While the measured $\rho(v_2^2,[\pt])$ increases monotonically toward lower multiplicity, the \ipgmusic calculation without IMA exhibits a weakly decreasing trend.
Furthermore, the sign of the \ipgmusic calculation without the IMA differs from that of the AMPT calculation, indicating that the initial shape-size correlations in the two models are opposite to each other. Overall, to improve the agreement between the IP-Glasma + MUSIC  +UrQMD calculations and the $\rho(v_2^2,[\pt])$ measurements in p--Pb collisions, the contribution from IMA is helpful, and better-tuned initial conditions with proper shape and size correlations are necessary.

Figure 4 compares the measured $\rho(v_2^2,[\pt])$ as a function of charged particle multiplicity in pp collisions with predictions from AMPT, \ipgmusic, and PYTHIA~8.
The PYTHIA~8 model predicts a positive correlation with only a weak \Nch dependence in pp collisions. 
This differs significantly from the $\rho(v_2^2,[\pt])$ measurements, which show a clear increasing trend toward lower \Nch.
Apparently, the measurements cannot be fully explained by a model without collective effects, which further confirms the existence of collective-like effects in the $\rho(v_2^2,[\pt])$ measurements.
A PYTHIA~8 calculation qualitatively describes the previous measurements~\cite{CMS:2024rvk}, which might be explained by the non-flow contamination originate from the usage of high-\pt particles~\cite{Zhang:2021phk}.
Meanwhile, both AMPT calculations and those from \ipgmusic present an increasing trend as the \Nch decreases. More specifically, the AMPT calculation shows a negative correlation between elliptic flow and mean transverse momentum in pp collisions. This is opposite to the AMPT calculations in Pb--Pb and p--Pb collisions, and also different from what is observed in the measurements in pp collisions. Besides, the \ipgmusic calculations in pp collisions are qualitatively similar to those in p--Pb collisions, with an increasing trend towards the low \Nch region. Both the default \ipgmusic calculations and the ones without IMA show negative $\rho(v_2^2,[\pt])$ results for $\Nch >10$, which is opposite to the measurements. Additionally, the \ipgmusic calculations without the IMA, unlike the p--Pb case, increase as multiplicity decreases in pp collisions. This shows different initial size and shape correlations in pp and p--Pb collisions in the IP-Glasma model. Moreover, both the calculations of \ipgmusic without the IMA and the one from AMPT show an opposite sign compared to the measurements, failing to provide a purely geometrical explanation for the results. This clearly highlights the current challenges of the AMPT model and the \ipgmusic model, particularly regarding the initial conditions, in describing the $\rho(v_2^2,[\pt])$ measurements in pp collisions.

\section{Summary}

This Letter presents the ALICE measurement of the correlations between elliptic flow ($v_{2}$) and mean transverse momentum ($[\pt]$) using the observable $\rho(v_2^2, [\pt])$, conducted for the first time in a small system with ALICE.
The measurements are presented as a function of charged particle multiplicity in pp, p--Pb, and Pb--Pb collisions. 
The $\rho(v_2^2, [p_\mathrm{T}])$ measurements show increasing trends in pp and p--Pb collisions as $\Nch$  decreases, and the results are similar across all three systems for $\Nch < 80$. 
Notably, the measurements in pp or p--Pb collisions cannot be reproduced by non-collective models such as PYTHIA~\cite{Skands:2014pea, Bierlich:2018xfw, Lim:2021auv}, revealing that the observed correlations cannot be attributed solely to non-flow effects and thus suggest a collective origin.
In Pb--Pb and p--Pb collisions, the AMPT transport model~\cite{Lin:2004en, Lim:2021auv} describes the measurements at low multiplicity but it
fails in pp collisions. The state-of-the-art hydrodynamics model \ipgmusic~\cite{Schenke:2012wb, Giacalone:2020byk}, with and without the initial momentum anisotropy (IMA) from Color Glass Condensate, qualitatively describes the measurements for $\Nch > 150$ in Pb--Pb collisions. However, these \ipgmusic calculations show significant deviations from the measurements for $\Nch < 150$ in Pb--Pb collisions, as well as in p--Pb and pp collisions. In some cases, negative correlations between mean transverse momentum and elliptic flow are predicted in the \ipgmusic calculations, opposite to what is observed in the measurements.
Additionally, the description of the measurements from \ipgmusic calculations without the IMA can be further improved by considering the IMA or by further tuning the initial size and shape correlations, such as those implemented in the AMPT model. 
The influence of the hydrodynamic starting time, $\tau_{\rm 0}$, is also significant, particularly in the low \Nch region.
A better understanding of the quantitative effects of this parameter on the magnitude of $\rho(v_2^2, [p_\mathrm{T}])$ is necessary.

Overall, the presented measurements on the correlations between flow and mean transverse momentum pose significant challenges to current state-of-the-art models and call for refined initial-state treatments. These new measurements offer new insights into the initial stage of collisions in small systems and also provide strong constraints on the existing theoretical models. This will significantly advance our understanding of the collective phenomena observed in small systems at the LHC.


\newenvironment{acknowledgement}{\relax}{\relax}
\begin{acknowledgement}
\section*{Acknowledgements}

The ALICE Collaboration would like to thank Ch. Shen, B. Schenke, and W. Zhao for providing the latest calculations from the state-of-the-art IP-Glasma+MUSIC+UrQMD models.


The ALICE Collaboration would like to thank all its engineers and technicians for their invaluable contributions to the construction of the experiment and the CERN accelerator teams for the outstanding performance of the LHC complex.
The ALICE Collaboration gratefully acknowledges the resources and support provided by all Grid centres and the Worldwide LHC Computing Grid (WLCG) collaboration.
The ALICE Collaboration acknowledges the following funding agencies for their support in building and running the ALICE detector:
A. I. Alikhanyan National Science Laboratory (Yerevan Physics Institute) Foundation (ANSL), State Committee of Science and World Federation of Scientists (WFS), Armenia;
Austrian Academy of Sciences, Austrian Science Fund (FWF): [M 2467-N36] and Nationalstiftung f\"{u}r Forschung, Technologie und Entwicklung, Austria;
Ministry of Communications and High Technologies, National Nuclear Research Center, Azerbaijan;
Rede Nacional de Física de Altas Energias (Renafae), Financiadora de Estudos e Projetos (Finep), Funda\c{c}\~{a}o de Amparo \`{a} Pesquisa do Estado de S\~{a}o Paulo (FAPESP) and The Sao Paulo Research Foundation  (FAPESP), Brazil;
Bulgarian Ministry of Education and Science, within the National Roadmap for Research Infrastructures 2020-2027 (object CERN), Bulgaria;
Ministry of Education of China (MOEC) , Ministry of Science \& Technology of China (MSTC) and National Natural Science Foundation of China (NSFC), China;
Ministry of Science and Education and Croatian Science Foundation, Croatia;
Centro de Aplicaciones Tecnol\'{o}gicas y Desarrollo Nuclear (CEADEN), Cubaenerg\'{\i}a, Cuba;
Ministry of Education, Youth and Sports of the Czech Republic, Czech Republic;
The Danish Council for Independent Research | Natural Sciences, the VILLUM FONDEN and Danish National Research Foundation (DNRF), Denmark;
Helsinki Institute of Physics (HIP), Finland;
Commissariat \`{a} l'Energie Atomique (CEA) and Institut National de Physique Nucl\'{e}aire et de Physique des Particules (IN2P3) and Centre National de la Recherche Scientifique (CNRS), France;
Bundesministerium f\"{u}r Forschung, Technologie und Raumfahrt (BMFTR) and GSI Helmholtzzentrum f\"{u}r Schwerionenforschung GmbH, Germany;
General Secretariat for Research and Technology, Ministry of Education, Research and Religions, Greece;
National Research, Development and Innovation Office, Hungary;
Department of Atomic Energy Government of India (DAE), Department of Science and Technology, Government of India (DST), University Grants Commission, Government of India (UGC) and Council of Scientific and Industrial Research (CSIR), India;
National Research and Innovation Agency - BRIN, Indonesia;
Istituto Nazionale di Fisica Nucleare (INFN), Italy;
Japanese Ministry of Education, Culture, Sports, Science and Technology (MEXT) and Japan Society for the Promotion of Science (JSPS) KAKENHI, Japan;
Consejo Nacional de Ciencia (CONACYT) y Tecnolog\'{i}a, through Fondo de Cooperaci\'{o}n Internacional en Ciencia y Tecnolog\'{i}a (FONCICYT) and Direcci\'{o}n General de Asuntos del Personal Academico (DGAPA), Mexico;
Nederlandse Organisatie voor Wetenschappelijk Onderzoek (NWO), Netherlands;
The Research Council of Norway, Norway;
Pontificia Universidad Cat\'{o}lica del Per\'{u}, Peru;
Ministry of Science and Higher Education, National Science Centre and WUT ID-UB, Poland;
Korea Institute of Science and Technology Information and National Research Foundation of Korea (NRF), Republic of Korea;
Ministry of Education and Scientific Research, Institute of Atomic Physics, Ministry of Research and Innovation and Institute of Atomic Physics and Universitatea Nationala de Stiinta si Tehnologie Politehnica Bucuresti, Romania;
Ministerstvo skolstva, vyskumu, vyvoja a mladeze SR, Slovakia;
National Research Foundation of South Africa, South Africa;
Swedish Research Council (VR) and Knut \& Alice Wallenberg Foundation (KAW), Sweden;
European Organization for Nuclear Research, Switzerland;
Suranaree University of Technology (SUT), National Science and Technology Development Agency (NSTDA) and National Science, Research and Innovation Fund (NSRF via PMU-B B05F650021), Thailand;
Turkish Energy, Nuclear and Mineral Research Agency (TENMAK), Turkey;
National Academy of  Sciences of Ukraine, Ukraine;
Science and Technology Facilities Council (STFC), United Kingdom;
National Science Foundation of the United States of America (NSF) and United States Department of Energy, Office of Nuclear Physics (DOE NP), United States of America.
In addition, individual groups or members have received support from:
Czech Science Foundation (grant no. 23-07499S), Czech Republic;
FORTE project, reg.\ no.\ CZ.02.01.01/00/22\_008/0004632, Czech Republic, co-funded by the European Union, Czech Republic;
European Research Council (grant no. 950692), European Union;
Deutsche Forschungs Gemeinschaft (DFG, German Research Foundation) ``Neutrinos and Dark Matter in Astro- and Particle Physics'' (grant no. SFB 1258), Germany;
FAIR - Future Artificial Intelligence Research, funded by the NextGenerationEU program (Italy).

\end{acknowledgement}

\bibliographystyle{utphys}   
\bibliography{bibliography}

\newpage
\appendix

\section{\texorpdfstring{$\rho(v_2^2,[\pt])$}{rho(v2,pt)} with different \texorpdfstring{$|\Delta\eta|$}{Delta eta} separations}
\label{app:gap}

\begin{figure}[!hbt]
\centering
    \includegraphics[width=0.6\textwidth]{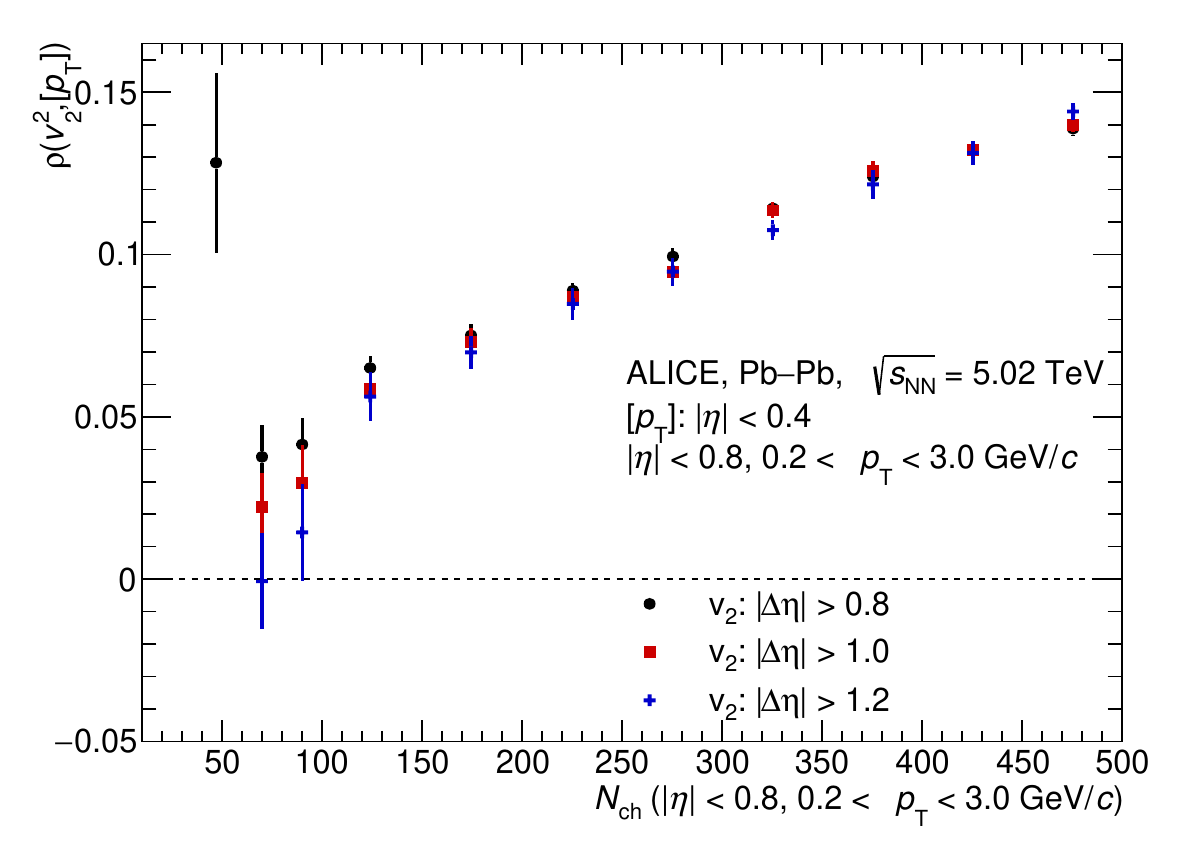}
    \caption{Measurements of $\rho(v_2^2, [\pt])$ as a function of charged-particle multiplicity at varying $|\Delta\eta|$ separations. The uncertainties shown are statistical.}
    \label{fig:rhovnpt_scan}
\end{figure}

The measurements of $\rho(v_2^2,[\pt])$ in Pb--Pb collisions, with varying $|\Delta\eta|$ separation in the  $v_2^2$ extraction, are shown in Fig.~\ref{fig:rhovnpt_scan}.
\clearpage 

\section{The ALICE Collaboration}
\label{app:collab}
\begin{flushleft} 
\small

I.J.~Abualrob\,\orcidlink{0009-0005-3519-5631}\,$^{\rm 113}$, 
S.~Acharya\,\orcidlink{0000-0002-9213-5329}\,$^{\rm 49}$, 
G.~Aglieri Rinella\,\orcidlink{0000-0002-9611-3696}\,$^{\rm 32}$, 
L.~Aglietta\,\orcidlink{0009-0003-0763-6802}\,$^{\rm 24}$, 
N.~Agrawal\,\orcidlink{0000-0003-0348-9836}\,$^{\rm 25}$, 
Z.~Ahammed\,\orcidlink{0000-0001-5241-7412}\,$^{\rm 133}$, 
S.~Ahmad\,\orcidlink{0000-0003-0497-5705}\,$^{\rm 15}$, 
I.~Ahuja\,\orcidlink{0000-0002-4417-1392}\,$^{\rm 36}$, 
Z.~Akbar$^{\rm 80}$, 
V.~Akishina\,\orcidlink{0009-0004-4802-2089}\,$^{\rm 38}$
% ...

\section*{Affiliation Notes}

$^{\rm I}$ Deceased\\
$^{\rm II}$ Also at: Max-Planck-Institut fur Physik, Munich, Germany\\
$^{\rm III}$ Also at: Czech Technical University in Prague (CZ)\\
$^{\rm IV}$ Also at: Instituto de Fisica da Universidade de Sao Paulo\\
$^{\rm V}$ Also at: Dipartimento DET del Politecnico di Torino, Turin, Italy\\
$^{\rm VI}$ Also at: Department of Applied Physics, Aligarh Muslim University, Aligarh, India\\
$^{\rm VII}$ Also at: Institute of Theoretical Physics, University of Wroclaw, Poland\\
$^{\rm VIII}$ Also at: Facultad de Ciencias, Universidad Nacional Aut\'{o}noma de M\'{e}xico, Mexico City, Mexico\\

\section*{Collaboration Institutes}

$^{1}$ A.I. Alikhanyan National Science Laboratory (Yerevan Physics Institute) Foundation, Yerevan, Armenia\\
$^{2}$ AGH University of Krakow, Cracow, Poland\\
$^{3}$ Bogolyubov Institute for Theoretical Physics, National Academy of Sciences of Ukraine, Kyiv, Ukraine\\
$^{4}$ Bose Institute, Department of Physics  and Centre for Astroparticle Physics and Space Science (CAPSS), Kolkata, India\\
$^{5}$ California Polytechnic State University, San Luis Obispo, California, United States\\
$^{6}$ Central China Normal University, Wuhan, China\\
$^{7}$ Centro de Aplicaciones Tecnol\'{o}gicas y Desarrollo Nuclear (CEADEN), Havana, Cuba\\
$^{8}$ Centro de Investigaci\'{o}n y de Estudios Avanzados (CINVESTAV), Mexico City and M\'{e}rida, Mexico\\
$^{9}$ Chicago State University, Chicago, Illinois, United States\\
$^{10}$ China Nuclear Data Center, China Institute of Atomic Energy, Beijing, China\\
$^{11}$ China University of Geosciences, Wuhan, China\\
$^{12}$ Chungbuk National University, Cheongju, Republic of Korea\\
$^{13}$ Comenius University Bratislava, Faculty of Mathematics, Physics and Informatics, Bratislava, Slovak Republic\\
$^{14}$ Creighton University, Omaha, Nebraska, United States\\
$^{15}$ Department of Physics, Aligarh Muslim University, Aligarh, India\\
$^{16}$ Department of Physics, Pusan National University, Pusan, Republic of Korea\\
$^{17}$ Department of Physics, Sejong University, Seoul, Republic of Korea\\
$^{18}$ Department of Physics, University of California, Berkeley, California, United States\\
$^{19}$ Department of Physics, University of Oslo, Oslo, Norway\\
$^{20}$ Department of Physics and Technology, University of Bergen, Bergen, Norway\\
$^{21}$ Dipartimento di Fisica, Universit\`{a} di Pavia, Pavia, Italy\\
$^{22}$ Dipartimento di Fisica dell'Universit\`{a} and Sezione INFN, Cagliari, Italy\\
$^{23}$ Dipartimento di Fisica dell'Universit\`{a} and Sezione INFN, Trieste, Italy\\
$^{24}$ Dipartimento di Fisica dell'Universit\`{a} and Sezione INFN, Turin, Italy\\
$^{25}$ Dipartimento di Fisica e Astronomia dell'Universit\`{a} and Sezione INFN, Bologna, Italy\\
$^{26}$ Dipartimento di Fisica e Astronomia dell'Universit\`{a} and Sezione INFN, Catania, Italy\\
$^{27}$ Dipartimento di Fisica e Astronomia dell'Universit\`{a} and Sezione INFN, Padova, Italy\\
$^{28}$ Dipartimento di Fisica `E.R.~Caianiello' dell'Universit\`{a} and Gruppo Collegato INFN, Salerno, Italy\\
$^{29}$ Dipartimento DISAT del Politecnico and Sezione INFN, Turin, Italy\\
$^{30}$ Dipartimento di Scienze MIFT, Universit\`{a} di Messina, Messina, Italy\\
$^{31}$ Dipartimento Interateneo di Fisica `M.~Merlin' and Sezione INFN, Bari, Italy\\
$^{32}$ European Organization for Nuclear Research (CERN), Geneva, Switzerland\\
$^{33}$ Faculty of Electrical Engineering, Mechanical Engineering and Naval Architecture, University of Split, Split, Croatia\\
$^{34}$ Faculty of Nuclear Sciences and Physical Engineering, Czech Technical University in Prague, Prague, Czech Republic\\
$^{35}$ Faculty of Physics, Sofia University, Sofia, Bulgaria\\
$^{36}$ Faculty of Science, P.J.~\v{S}af\'{a}rik University, Ko\v{s}ice, Slovak Republic\\
$^{37}$ Faculty of Technology, Environmental and Social Sciences, Bergen, Norway\\
$^{38}$ Frankfurt Institute for Advanced Studies, Johann Wolfgang Goethe-Universit\"{a}t Frankfurt, Frankfurt, Germany\\
$^{39}$ Fudan University, Shanghai, China\\
$^{40}$ Gauhati University, Department of Physics, Guwahati, India\\
$^{41}$ Helmholtz-Institut f\"{u}r Strahlen- und Kernphysik, Rheinische Friedrich-Wilhelms-Universit\"{a}t Bonn, Bonn, Germany\\
$^{42}$ Helsinki Institute of Physics (HIP), Helsinki, Finland\\
$^{43}$ High Energy Physics Group,  Universidad Aut\'{o}noma de Puebla, Puebla, Mexico\\
$^{44}$ Horia Hulubei National Institute of Physics and Nuclear Engineering, Bucharest, Romania\\
$^{45}$ HUN-REN Wigner Research Centre for Physics, Budapest, Hungary\\
$^{46}$ Indian Institute of Technology Bombay (IIT), Mumbai, India\\
$^{47}$ Indian Institute of Technology Indore, Indore, India\\
$^{48}$ INFN, Laboratori Nazionali di Frascati, Frascati, Italy\\
$^{49}$ INFN, Sezione di Bari, Bari, Italy\\
$^{50}$ INFN, Sezione di Bologna, Bologna, Italy\\
$^{51}$ INFN, Sezione di Cagliari, Cagliari, Italy\\
$^{52}$ INFN, Sezione di Catania, Catania, Italy\\
$^{53}$ INFN, Sezione di Padova, Padova, Italy\\
$^{54}$ INFN, Sezione di Pavia, Pavia, Italy\\
$^{55}$ INFN, Sezione di Torino, Turin, Italy\\
$^{56}$ INFN, Sezione di Trieste, Trieste, Italy\\
$^{57}$ Inha University, Incheon, Republic of Korea\\
$^{58}$ Institute for Gravitational and Subatomic Physics (GRASP), Utrecht University/Nikhef, Utrecht, Netherlands\\
$^{59}$ Institute of Experimental Physics, Slovak Academy of Sciences, Ko\v{s}ice, Slovak Republic\\
$^{60}$ Institute of Physics, Homi Bhabha National Institute, Bhubaneswar, India\\
$^{61}$ Institute of Physics of the Czech Academy of Sciences, Prague, Czech Republic\\
$^{62}$ Institute of Space Science (ISS), Bucharest, Romania\\
$^{63}$ Institut f\"{u}r Kernphysik, Johann Wolfgang Goethe-Universit\"{a}t Frankfurt, Frankfurt, Germany\\
$^{64}$ Instituto de Ciencias Nucleares, Universidad Nacional Aut\'{o}noma de M\'{e}xico, Mexico City, Mexico\\
$^{65}$ Instituto de F\'{i}sica, Universidade Federal do Rio Grande do Sul (UFRGS), Porto Alegre, Brazil\\
$^{66}$ Instituto de F\'{\i}sica, Universidad Nacional Aut\'{o}noma de M\'{e}xico, Mexico City, Mexico\\
$^{67}$ iThemba LABS, National Research Foundation, Somerset West, South Africa\\
$^{68}$ Jeonbuk National University, Jeonju, Republic of Korea\\
$^{69}$ Korea Institute of Science and Technology Information, Daejeon, Republic of Korea\\
$^{70}$ Laboratoire de Physique Subatomique et de Cosmologie, Universit\'{e} Grenoble-Alpes, CNRS-IN2P3, Grenoble, France\\
$^{71}$ Lawrence Berkeley National Laboratory, Berkeley, California, United States\\
$^{72}$ Lund University Department of Physics, Division of Particle Physics, Lund, Sweden\\
$^{73}$ Marietta Blau Institute, Vienna, Austria\\
$^{74}$ Nagasaki Institute of Applied Science, Nagasaki, Japan\\
$^{75}$ Nara Women{'}s University (NWU), Nara, Japan\\
$^{76}$ National and Kapodistrian University of Athens, School of Science, Department of Physics , Athens, Greece\\
$^{77}$ National Centre for Nuclear Research, Warsaw, Poland\\
$^{78}$ National Institute of Science Education and Research, Homi Bhabha National Institute, Jatni, India\\
$^{79}$ National Nuclear Research Center, Baku, Azerbaijan\\
$^{80}$ National Research and Innovation Agency - BRIN, Jakarta, Indonesia\\
$^{81}$ Niels Bohr Institute, University of Copenhagen, Copenhagen, Denmark\\
$^{82}$ Nikhef, National institute for subatomic physics, Amsterdam, Netherlands\\
$^{83}$ Nuclear Physics Group, STFC Daresbury Laboratory, Daresbury, United Kingdom\\
$^{84}$ Nuclear Physics Institute of the Czech Academy of Sciences, Husinec-\v{R}e\v{z}, Czech Republic\\
$^{85}$ Oak Ridge National Laboratory, Oak Ridge, Tennessee, United States\\
$^{86}$ Ohio State University, Columbus, Ohio, United States\\
$^{87}$ Physics department, Faculty of science, University of Zagreb, Zagreb, Croatia\\
$^{88}$ Physics Department, Panjab University, Chandigarh, India\\
$^{89}$ Physics Department, University of Jammu, Jammu, India\\
$^{90}$ Physics Program and International Institute for Sustainability with Knotted Chiral Meta Matter (WPI-SKCM$^{2}$), Hiroshima University, Hiroshima, Japan\\
$^{91}$ Physikalisches Institut, Eberhard-Karls-Universit\"{a}t T\"{u}bingen, T\"{u}bingen, Germany\\
$^{92}$ Physikalisches Institut, Ruprecht-Karls-Universit\"{a}t Heidelberg, Heidelberg, Germany\\
$^{93}$ Physik Department, Technische Universit\"{a}t M\"{u}nchen, Munich, Germany\\
$^{94}$ Politecnico di Bari and Sezione INFN, Bari, Italy\\
$^{95}$ Research Division and ExtreMe Matter Institute EMMI, GSI Helmholtzzentrum f\"ur Schwerionenforschung GmbH, Darmstadt, Germany\\
$^{96}$ Saga University, Saga, Japan\\
$^{97}$ Saha Institute of Nuclear Physics, Homi Bhabha National Institute, Kolkata, India\\
$^{98}$ School of Physics and Astronomy, University of Birmingham, Birmingham, United Kingdom\\
$^{99}$ Secci\'{o}n F\'{\i}sica, Departamento de Ciencias, Pontificia Universidad Cat\'{o}lica del Per\'{u}, Lima, Peru\\
$^{100}$ SUBATECH, IMT Atlantique, Nantes Universit\'{e}, CNRS-IN2P3, Nantes, France\\
$^{101}$ Sungkyunkwan University, Suwon City, Republic of Korea\\
$^{102}$ Suranaree University of Technology, Nakhon Ratchasima, Thailand\\
$^{103}$ Technical University of Ko\v{s}ice, Ko\v{s}ice, Slovak Republic\\
$^{104}$ The Henryk Niewodniczanski Institute of Nuclear Physics, Polish Academy of Sciences, Cracow, Poland\\
$^{105}$ The University of Texas at Austin, Austin, Texas, United States\\
$^{106}$ Universidad Aut\'{o}noma de Sinaloa, Culiac\'{a}n, Mexico\\
$^{107}$ Universidade de S\~{a}o Paulo (USP), S\~{a}o Paulo, Brazil\\
$^{108}$ Universidade Estadual de Campinas (UNICAMP), Campinas, Brazil\\
$^{109}$ Universidade Federal do ABC, Santo Andre, Brazil\\
$^{110}$ Universitatea Nationala de Stiinta si Tehnologie Politehnica Bucuresti, Bucharest, Romania\\
$^{111}$ University of Cape Town, Cape Town, South Africa\\
$^{112}$ University of Derby, Derby, United Kingdom\\
$^{113}$ University of Houston, Houston, Texas, United States\\
$^{114}$ University of Jyv\"{a}skyl\"{a}, Jyv\"{a}skyl\"{a}, Finland\\
$^{115}$ University of Kansas, Lawrence, Kansas, United States\\
$^{116}$ University of Liverpool, Liverpool, United Kingdom\\
$^{117}$ University of Science and Technology of China, Hefei, China\\
$^{118}$ University of Silesia in Katowice, Katowice, Poland\\
$^{119}$ University of South-Eastern Norway, Kongsberg, Norway\\
$^{120}$ University of Tennessee, Knoxville, Tennessee, United States\\
$^{121}$ University of the Witwatersrand, Johannesburg, South Africa\\
$^{122}$ University of Tokyo, Tokyo, Japan\\
$^{123}$ University of Tsukuba, Tsukuba, Japan\\
$^{124}$ Universit\"{a}t M\"{u}nster, Institut f\"{u}r Kernphysik, M\"{u}nster, Germany\\
$^{125}$ Universit\'{e} Clermont Auvergne, CNRS/IN2P3, LPC, Clermont-Ferrand, France\\
$^{126}$ Universit\'{e} de Lyon, CNRS/IN2P3, Institut de Physique des 2 Infinis de Lyon, Lyon, France\\
$^{127}$ Universit\'{e} de Strasbourg, CNRS, IPHC UMR 7178, F-67000 Strasbourg, France, Strasbourg, France\\
$^{128}$ Universit\'{e} Paris-Saclay, Centre d'Etudes de Saclay (CEA), IRFU, D\'{e}partment de Physique Nucl\'{e}aire (DPhN), Saclay, France\\
$^{129}$ Universit\'{e}  Paris-Saclay, CNRS/IN2P3, IJCLab, Orsay, France\\
$^{130}$ Universit\`{a} degli Studi di Foggia, Foggia, Italy\\
$^{131}$ Universit\`{a} del Piemonte Orientale, Vercelli, Italy\\
$^{132}$ Universit\`{a} di Brescia, Brescia, Italy\\
$^{133}$ Variable Energy Cyclotron Centre, Homi Bhabha National Institute, Kolkata, India\\
$^{134}$ Warsaw University of Technology, Warsaw, Poland\\
$^{135}$ Wayne State University, Detroit, Michigan, United States\\
$^{136}$ Yale University, New Haven, Connecticut, United States\\
$^{137}$ Yildiz Technical University, Istanbul, Turkey\\
$^{138}$ Yonsei University, Seoul, Republic of Korea\\
$^{139}$ Affiliated with an institute formerly covered by a cooperation agreement with CERN\\
$^{140}$ Affiliated with an international laboratory covered by a cooperation agreement with CERN.\\

\end{flushleft} 

\end{document}